# Tunneling photo-thermoelectric effect in monolayer graphene/bilayer hexagonal boron nitride/bilayer graphene asymmetric van der Waals tunnel junctions

Sabin Park[1], Rai Moriya[1,*], Yijin Zhang[1], Kenji Watanabe[2], Takashi Taniguchi[3], and Tomoki Machida[1,*]

[1] *Institute of Industrial Science, University of Tokyo, 4-6-1 Komaba, Meguro, Tokyo 153-8505, Japan*

[2] *Research Center for Electronic and Optical Materials, National Institute for Materials Science, 1-1 Namiki, Tsukuba 305-0044, Japan*

[3] *Research Center for Materials Nanoarchitectonics, National Institute for Materials Science, 1-1 Namiki, Tsukuba 305-0044, Japan*

**Graphene is known to exhibit a pronounced photo-thermoelectric effect (PTE) in its in-plane carrier transport and attracting attention toward various optoelectronic applications. In this study, we demonstrate an out-of-plane PTE by utilizing electron tunneling across a barrier, namely, the tunneling photo-thermoelectric effect (TPTE). This was achieved in a monolayer graphene (MLG)/bilayer hexagonal boron nitride (*h*-BN)/bilayer graphene (BLG) asymmetric tunnel junction. MLG and BLG exhibit different cyclotron resonance (CR) optical absorption energies when their energies are Landau quantized under an out-of-plane magnetic field. We tuned the magnetic field under mid-infrared (MIR) irradiation to bring MLG into CR conditions, whereas BLG was not in CR. The CR absorption in the MLG generates an electron temperature difference between the MLG and BLG, and induces an out-of-plane**

**TPTE voltage across the *h*-BN tunnel barrier. The TPTE exhibited a unique dependence on the Fermi energy of the MLG, which differed from that of the in-plane PTE of the MLG. The TPTE signal was large when the Fermi energy of the MLG was tuned near the phase transition between the quantum Hall state (QHS) and non-QHS, that is, the transition between carrier localization and delocalization. The TPTE provides another degree of freedom for probing the electronic and optoelectronic properties of two-dimensional material heterostructures.**

----------

*E-mail: moriyar@iis.u-tokyo.ac.jp; tmachida@iis.u-tokyo.ac.jp

Graphene is known to exhibit a significant photo-thermoelectric effect (PTE) for generating photovoltage or photocurrent in a broad wavelength range from visible to millimeter wave [1-3]. Thus, PTE in graphene and its heterostructures with other two-dimensional (2D) materials have received considerable attention for optoelectronic device applications. Recently, PTE has received more attention for its use in probing various quasiparticle states of graphene, such as plasmon [4] and polariton [5], as well as the local electric structure of moiré superlattices in twisted bilayer graphene [6]. The development of PTE of graphene is in high demand for further progress in 2D material research. So far, most studies on PTE have utilized the in-plane carrier transport of graphene [7-12]. It is known that the detection of PTE in in-plane graphene devices requires asymmetric device structures, and which impose serious restrictions on device fabrication [12-15]. We provide a detailed discussion of this using a schematic illustration of the two-terminal graphene device shown in Supplementary Figs. S1(a) and S1(b). Under irradiation, the optical absorption of graphene immediately increases its electron temperature. The temperature difference between graphene and the electrode, $\Delta T$, generates a thermoelectric voltage $V_{ph} \propto S\Delta T$ at each of the electrode/graphene junctions, where $S$ is the thermoelectric coefficient of graphene. In an ideal graphene device with a symmetric structure, $V_{ph}$ from the respective junctions have the same amplitude but opposite signs to cancel each other. Asymmetric device structures, for example, asymmetric carrier density between two electrode/graphene junctions using local gating, are necessary to detect the PTE signal [10,12,16,17].

In this article, we propose a novel type of PTE that utilizes vertical transport through a tunnel junction. In tunnel junctions with asymmetric electrodes of monolayer graphene (MLG) and bilayer graphene (BLG), asymmetric light absorption can be realized by choosing a condition in which only MLG exhibits light absorption [Supplementary Fig. S1(c)]. Consequently, a temperature gradient is generated between the MLG and BLG across the tunnel barrier. This subsequently induces PTE in the tunnel junction. A more specific device structure is presented in Fig. 1(a). A vertical tunnel junction of MLG/bilayer $h$-BN/BLG was fabricated, where the MLG and BLG were used as the bottom and top electrodes, respectively. Under the application of a perpendicular magnetic field $B$, discrete energy levels, so-called Landau levels, are formed in the MLG and BLG, as illustrated in Figs. 1(b) and 1(c). Because the Landau-quantized energies of MLG and BLG are different, their optical absorptions between the Landau levels, known as cyclotron resonance (CR), occur at different energies. We tuned the magnetic field under infrared irradiation to bring the MLG into CR conditions, while the BLG was not in CR, to induce electron heating only in the MLG. Consequently, a temperature gradient $\Delta T$ between the MLG and BLG was generated, as illustrated in Fig. 1(d). This subsequently generates $V_{\text{ph}}$ across the $h$-BN tunnel barrier. This phenomenon can be regarded as a tunnelling photo-thermoelectric effect (TPTE). TPTE is different from conventional in-plane PTE in several ways: (1) the TPTE voltage is generated in the out-of-plane direction across a thin tunnel barrier, (2) the asymmetric electrodes of MLG and BLG greatly simplify the device structure, and (3) TPTE is sensitive to the electronic states of the electrode material owing to the nature of tunneling. These unique characteristics of TPTE provide a novel method for expanding PTE in graphene for a wide variety of applications. In this study, we demonstrate TPTE in an MLG/$h$-BN/BLG asymmetric tunnel junction.

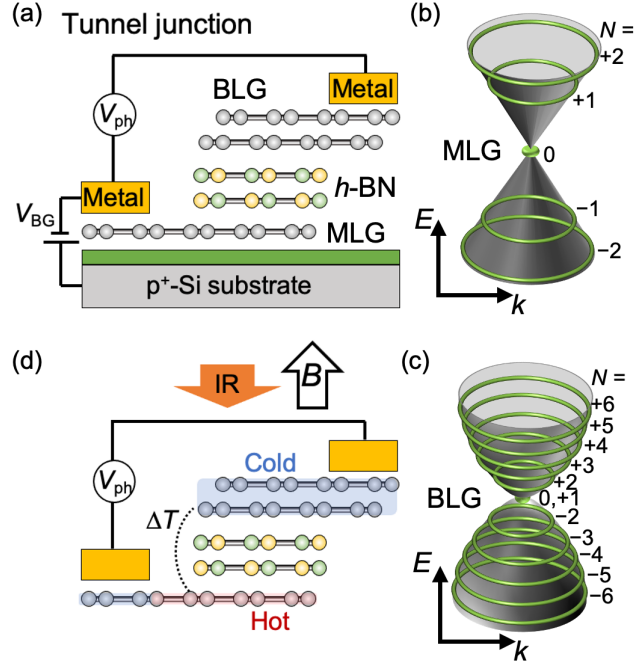

**Figure 1: (a) Schematic illustration of the device structure of the MLG/*h*-BN/BLG tunnel junction. (b,c) Illustration of Landau-quantized bands of (b) MLG and (c) BLG. The Landau level index (*N*) also shown. (d) Under IR irradiation in magnetic field *B*, the cyclotron resonance (CR) of the MLG increases its electron temperature, creating an out-of-plane temperature gradient Δ*T* between the MLG region and BLG.**

A MLG/bilayer *h*-BN/BLG tunnel junction illustrated in Fig. 1(a) was fabricated with van der Waals (vdW) pick-up method [18-21]. Graphene and *h*-BN were exfoliated from their bulk crystals and deposited on a 290-nm-thick $SiO_2$/Si substrate prior to device fabrication. We selected 30 − 50 nm thick *h*-BN for device encapsulation. The fabricated device was placed on a 290-nm-thick $SiO_2$/p-doped Si substrate. During the measurement, a back-gate voltage $V_{BG}$ was applied to the doped Si substrate to control the carrier density in the MLG. All measurements were performed in a variable-temperature cryostat (temperature range 2 − 300 K) equipped with a superconducting magnet (magnetic field *B* from 0 to 15 T). Resistance-area product $RA$ of the junction was determined as $RA \sim 7 \times 10^5$ $\Omega \cdot \mu m^2$ at $V_{BG} = 0$ V and $T = 3.0$ K. By using a lock-in amplifier (NF Corp. LI5640), the photovoltage $V_{ph} = V(\text{irradiation}) - V(\text{dark})$ was measured under infrared (IR) irradiation using wavelength-tunable $CO_2$ laser (Access Laser Inc. AL20G, wavelength $\lambda =$ 9.24–10.675 μm). The irradiation was modulated using an optical chopper with an operating frequency of ~11 Hz. Light from the laser was guided to the sample inside the cryostat using an optical fiber and light pipe, enabling photovoltage measurements. The power density of the laser was estimated as $3.8 \times 10^{-2}$ W/cm$^2$ at the sample position. The differential conductance, $dI/dV$, was measured using a function generator (NF Corp. WF1974) to generate ac bias voltage $V = 0.5$ mV superimposed on a dc bias voltage $V_{dc}$, and tunnel current $I$ flowing through the junction was measured with a current amplifier (DL Instruments, 1211) and a lock-in amplifier.

The conductance of the tunnel junction in the out-of-plane direction, $\sigma_T$, was determined by measuring $dI/dV$ between the metal electrodes on the MLG and BLG, as illustrated in Fig. 1(a). The $\sigma_T$ versus $V_{BG}$ measured for different $B$ values at $T = 3.0$ K and $V_{dc} = 40$ mV are presented in Fig. 2(a). Here, $B$ was changed from $B = 0$ T (bottom trace) to 12.7 T (top trace) and the $V_{BG}$ value with respect to that for the charge neutrality point (CNP) of the MLG, $V_{CNP}$, is plotted on the x-axis. The results show the development of quantum Hall states from low to high $B$. At $B = 0$ T, the $\sigma_T$ is minimized at the CNP ($V_{BG} - V_{CNP} = 0$ V). With increasing $B$, multiple minima in $\sigma_T$ develop. This phenomenon is due to the Landau quantization of the MLG. In the high $B$, these minima become wider and show plateau features with nearly zero conductance, which corresponds to QHS. The red dashed line in the figure indicates the Landau-level (LL) filling factor $v = hn/eB = \pm 2$ of the MLG, where $h$ is the Planck constant, $n$ is the carrier density of the MLG, and $e$ is the elementary charge. The influence of the QHS of the BLG is not apparent in Fig. 2(a). We believe that the influence of $V_{BG}$ on the BLG is much smaller than that of MLG because of the screening of the back-gate electric field by the MLG. Thus, the carrier density of BLG remained close to zero throughout our experiment. Next, an image plot of photovoltage $V_{ph}$ as a function of $V_{BG}$ and $B$ at $T \sim 3.0$ K and $\lambda = 9.569$ μm is presented in Fig. 2(b). The intense $V_{ph}$ signals appeared at around $B = 10.7$ and $11.5$ T and its sign changes with $V_{BG}$. The same measurements were performed with irradiation wavelengths of $\lambda = 9.250$ μm, $10.247$ μm, and $10.611$ μm. The relationship between $|V_{ph}|$ and $B$ for fixed $V_{BG}$ values is shown in Fig. 2(c). Here, the $V_{BG}$ values for each trace were chosen for the most intense $V_{ph}$ signal [e.g., $V_{BG} \sim -36$ V for Fig. 2(b)]. $|V_{ph}|$ exhibited double peaks around $B = 8 - 13$ T and their peak positions changed with $\lambda$. We found that the $B$ and $\lambda$ values to observe the peak of $|V_{ph}|$ follow the condition for CR absorption in the MLG. Here, a relationship between the energy of irradiated $\lambda$ of 134.2 meV (9.24 μm), 129.6 meV (9.569 μm), 121.0 meV (10.247 μm), and 116.8 meV (10.611 μm) versus the peak $B$ values for the double peak structures in Fig. 2(c) was plotted in Fig. 2(d) with black solid circle and open square. An analytical formula of the CR transition energy for the $T_1$ transition, $E_{T1}$, can be written as follows [22-24];

$$E_{T1} = \sqrt{2e\hbar v_F^2 B + \left(\frac{E_g}{2}\right)^2} \pm \frac{E_g}{2}, \quad (1)$$

where $\hbar$ denotes the reduced Planck constant, $v_F$ is the Fermi velocity of graphene, and $E_g$ is the bandgap of graphene. In our device, the interaction between MLG and adjacent h-BN opens the band gap in MLG, splitting $N = 0$ LL into $N = 0_+$ and $0_-$ LLs [22,24].

Consequently, the $T_1$ transition energy is also split with a difference equal to $E_g$. The ± sign in the Eq. (1) denotes the two different $T_1$ transitions. We plot the eq. (1) in Fig. 2(e) as red (+ sign) and blue (− sign) dashed lines using the parameters of $v_F$ =1.05 × 10$^6$ m/s and $E_g$ = 4.27 meV. These values were within the range typically observed for *h*-BN-encapsulated MLG [10,25-27]. We obtained good agreement between the experiment and Eq. (1) [A detailed comparison of the experimental data and Eq. (1) with different parameters $v_F$ and $E_g$, are shown in Figs. S3(a) and S3(b).] From these results, we demonstrate that the observed peaks of $V_{ph}$ satisfy the condition that only the MLG is in CR. This is a prerequisite for observing the TPTE.

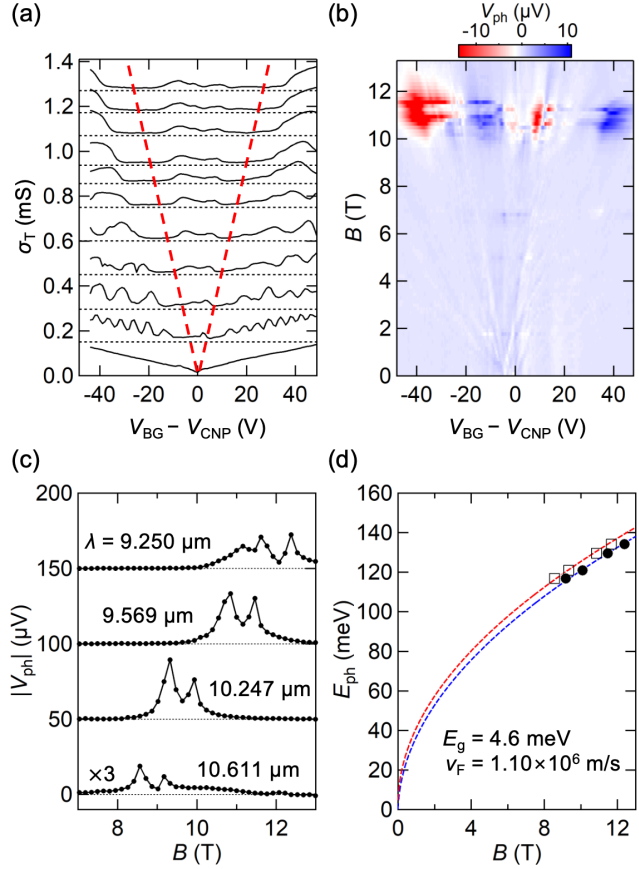

**Figure 2:** (a) A conductance of the tunnel junction $\sigma_T$ versus $V_{BG}$ for different $B$ values measured at $T = 3.0$ K under application of constant dc bias voltage of $V_{SD} = 40$ mV. The data are offset for clarity (the offsets are indicated by the horizontal dashed lines). The $B$ values for each date are 12.70, 11.72, 10.70, 9.38, 8.56, 7.50, 6.01, 4.50, 2.96, 1.50, and 0.00 T (from top to bottom). (b) A photovoltage $V_{ph}$ versus $V_{BG}$ at different $B$ measured at $\lambda = 9.657$ μm and $T = 3.0$ K. (c) The amplitude of $V_{ph}$ versus $B$ measured at different $\lambda$ and $T = 3.0$ K. The data were offset for clarity (the offsets are indicated by the horizontal dotted lines). (d) The solid circle and open square indicate a relationship between the energy of irradiation and the $B$ value of the $|V_{ph}|$ peaks in panel (c). A CR energy calculated with Fermi velocity $v_F = 1.05 \times 10^6$ m/s and gap $E_g = 4.27$ meV are plotted as red and blue dashed lines.

Next, we discuss the details of the $V_{ph}$ signals under the CR resonance conditions of the MLG based on the TPTE. First, the photovoltage induced by PTE generally follows the relation $V_{ph} \propto (1/\sigma)d\sigma/dE$, where $\sigma$ is the conductance and $E$ is the energy of the carrier [9,28]. This implies that a large PTE photovoltage is expected for smaller $\sigma$ and larger $d\sigma/dE$. The graphene-based tunnel junction near the transition between the QHS and non-QHS satisfies this condition. When graphene is in a QHS and non-QHS, its electrons are localized and delocalized, respectively. Because localized electrons do not contribute to the conductance, we believe that the tunneling conductance $\sigma_T$ is small when graphene is in the QHS. When the electronic state of graphene changes from QHS to non-QHS by tuning its $E$, $\sigma_T$ significantly increases because the delocalized electrons contribute significantly to tunneling; thus, a large $d\sigma_T/dE$ is expected. Therefore, we expect the emergence of a large $(1/\sigma_T)d\sigma_T/dE$, and thus a large $V_{ph}$ in the vicinity of the localization-delocalization (QHS to non-QHS) transition in the graphene electrode of the tunnel junction. A more detailed discussion is provided in the supplementary information Fig. S4. To confirm this hypothesis, the experimental data for $V_{ph}$ are compared with the behavior expected for TPTE in the next paragraph.

In Fig. 3(a), the experimental data of the tunnel conductance $\sigma_T$ versus $V_{BG}$ at a magnetic field of $B = 10.7$ T (at one of the CR conditions) are presented. The two minima in $\sigma_T$ shaded in solid light blue correspond to two different QHS with a filling factor of $\nu = \pm 2$. The derivative of this experimental data, $\sigma_T' = \partial \sigma_T / \partial V_{BG}$ is shown in Fig. 3(b). The PTE in the tunnel junction, thus the TPTE voltage, can be written as $V_{ph} = S_T \Delta T$, where $S_T$ denotes the thermoelectric coefficient of the tunnel junction. Thus, $S_T$ determines the shape of $V_{ph}$. According to Mott formula, $S_T \propto \mp (d\sigma_T/dE)/\sigma_T \approx \mp (d\sigma_T/dV_{BG})/\sigma_T = \mp \sigma_T'/\sigma_T$, where $\mp$ sign is chosen whether electron or hole is a transport carrier. We obtained $+\sigma_T'/\sigma_T$ directly from the experimental data of $\sigma_T$ [Fig. 3(a)] and its derivative $\sigma_T'$ [Fig. 3(b)], as presented in Fig. 3(c). This should be compared with the measured $V_{ph}$ versus $V_{BG}$ at $B =$ 10.6 T in Fig. 3(d). The $V_{BG}$ dependences of $+\sigma_T'/\sigma_T$ [Fig. 3(c)] and $V_{ph}$ [Fig. 3(d)] demonstrated good agreement. Both datasets exhibited peaks and dips on both sides of the QHS (solid light blue squares in Fig. 3, corresponding to a QHS of $\nu = \pm 2$.). The + sign in front of $\sigma_T'/\sigma_T$ was chosen to obtain consistency in the shape between $\sigma_T'/\sigma_T$ and $V_{ph}$. This suggests that the carriers contributing to the tunnelling are holes. This is also consistent with the reported characteristics of the $h$-BN barrier such that the tunnel barrier height for holes is lower than that of electrons [29,30]. The $V_{BG}$ positions of the peaks and dips appear when the slope of $\sigma_T$ versus $V_{BG}$ is large, which is the boundary between the QHS and non-QHS, as assigned in Fig. 3(a). In other words, the peaks and dips appear at the localization-delocalization transition, supporting our hypothesis of PTE enhancement near the transition. This result confirmed that the observed photovoltage was due to the CR absorption-induced thermoelectric voltage at the tunnel junction. The $\Delta T$ between the MLG and BLG induced by the CR of the MLG generated $V_{ph}$ near the localization-delocalization transition. The actual $\Delta T$ on our device was estimated to be $\Delta T \sim 8$ K from its in-plane transport measurements (see Supplementary Fig. S5). The observed phenomena shown here satisfy the condition of the TPTE, which has the following unique characteristics: (1) the TPTE voltage is generated in the out-of-plane direction across the tunnel barriers, (2) the asymmetric electrode structure of MLG and BLG realizes the generation of a temperature difference between the electrodes, and (3) the TPTE is enhanced in the vicinity of the localization-delocalization transition. In addition, the observed character is different from the in-plane PTE in Landau-quantized MLG. In Landau-quantized MLG, the PTE signal is large when its $E_F$ is located near the CNP and significantly decreases when $E_F$ is away from the CNP. This is due to the significant influence of Hall conductivity [9,10,12,23,31]. The TPTE signal did not appear at the CNP as shown in Fig. 3, because the Hall conductivity contribution was absent in the tunnel junction. Instead, as we see above, the TPTE exhibits signal maxima and minima away from the CNP; TPTE signals are large at the boundary between the QHS and non-QHS. These features appeal to the unique characteristics of TPTE in tunnel junctions.

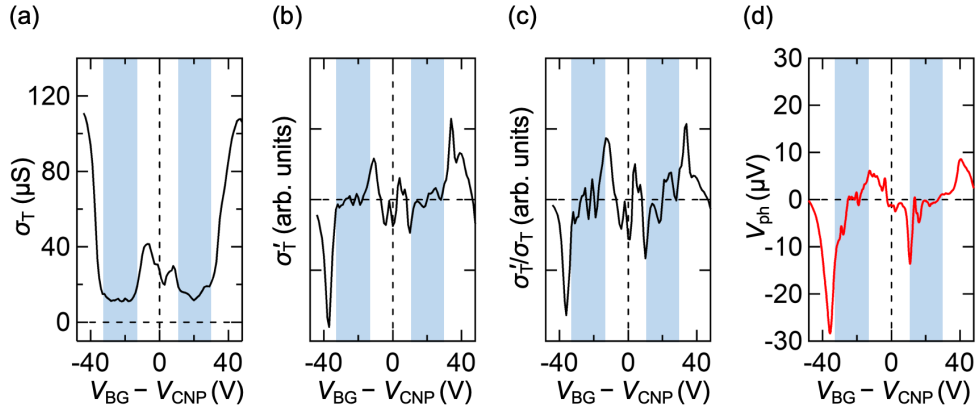

**Figure 3: (a) A conductance of the tunnel junction $\sigma_T$ versus $V_{BG}$ measured at $V_{SD}$ = 40 mV, $B$ = 10.7 T, and $T$ = 3.0 K. (b) Derivative of panel (a). (c) +$\sigma_T'(V_{BG})/\sigma_T(V_{BG})$ versus $V_{BG}$ calculated from the data in panel (a). (d) A photovoltage $V_{ph}$ versus $V_{BG}$ measured at $B$ = 10.7 T and $T$ = 3.0 K under the irradiation of $\lambda$ = 9.657 μm. Light blue square in panels (a) to (d) indicates positions of $v$ = ±2 QHS.**

As presented in the previous paragraph, the presence of a sharp transition between the QHS and non-QHS is crucial for observing the TPTE in our device. Because the QHS can be disturbed by increasing the temperature, we carried out temperature dependence of the $V_{ph}$ data to reveal the importance of the QHS. Fig. 4 plots the temperature dependence of the maxima (minima) of $V_{ph}$ under one of the CR conditions in the MLG. An image plot of $V_{ph}$ as a function of $V_{BG}$ and $T$ is shown in the inset of Fig. 4. Two maxima of $V_{ph}$ at $V_{BG}$

= –10 and +45 V and two minima at $V_{BG}$ = –35 and +12 V were identified at the lowest temperatures. The $V_{ph}$ amplitudes of these maxima (minima) as a function of temperature are plotted in Fig. 4. These values decrease with increasing temperature and diminish at $T$ ~ 110 K. In terms of the transport properties of graphene, the localized QHS of MLG gradually diminishes and becomes delocalized non-QHS with increasing temperature owing to the reduction in the inelastic-scattering length. The enhancement of $\sigma_T'/\sigma_T$ due to the localization to delocalization transition also decreases with increasing temperature. Therefore, the presence of a QHS in the MLG is crucial for observing the TPTE in our device.

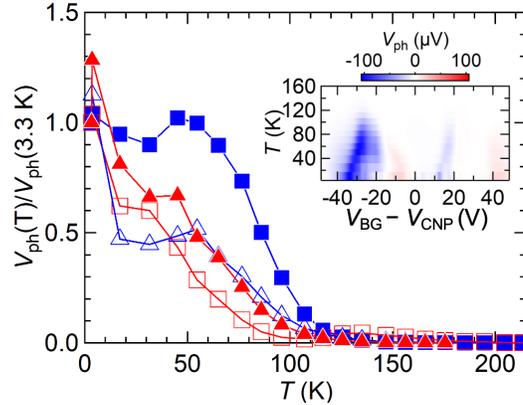

**Figure 4: A temperature dependence of the amplitude of $V_{ph}$ measured at $B$ = 10.6 T and $\lambda$ = 9.657 μm. $V_{ph}$ value was normalized by its value at $T$ = 3.3 K. Two maxima of $V_{ph}$ at $V_{BG}$ = –10 and +45 V and two minima at $V_{BG}$ = –35 and +12 V were identified at the $T$ = 3.3 K and their temperature dependence are plotted. Inset: image plot of $V_{ph}$ as a function of $V_{BG}$ and $T$.**

In summary, we demonstrated out-of-plane TPTE in an MLG/bilayer $h$-BN/BLG asymmetric tunnel junction. By tuning the magnetic field under mid-infrared (MIR) irradiation to induce CR absorption only in the MLG, an electron temperature difference between the MLG and BLG was generated, which induced an out-of-plane TPTE voltage across the *h*-BN tunnel barrier. The TPTE is significantly different from the conventional in-plane PTE and provides another degree of freedom for probing the electronic and optoelectronic properties of two-dimensional material heterostructures.

## Supplementary materials

See the supplementary material for supplementary figures from Fig. S1 to S5.

## Acknowledgements

This work was supported by JST-CREST, JST-Mirai, and JST-PRESTO (Grant Numbers JPMJCR15F3, JPMJCR20B4, JPMJMI21G9, JPMJPR20L5); JSPS KAKENHI (Grant Numbers JP20H00127, JP20H00354, JP21H05232, JP21H05233, JP21H05234, JP22H01898, JP22K18317, JP23H02052, JP24K01293); World Premier International Research Center Initiative (WPI), MEXT, Japan; World-leading Innovative Graduate Study Program for Materials Research, Information, and Technology (MERIT-WINGS) of the University of Tokyo.

## Conflict of Interest

The authors have no conflicts to disclose.

## Author Contributions

**Sabin Park:** Data curation (equal); Formal analysis (equal); Funding acquisition (equal); Investigation (equal); Resources (equal); Software (equal); Validation (equal);

Visualization (equal); Writing – original draft (equal); Writing – review & editing (equal).

**Rai Moriya:** Conceptualization (equal); Data curation (equal); Formal analysis (equal); Funding acquisition (equal); Investigation (equal); Methodology (equal); Project administration (equal); Resources (equal); Software (equal); Validation (equal); Visualization (equal); Writing – original draft (equal); Writing – review & editing (equal).

**Yijin Zhang:** Funding acquisition (equal); Resources (equal).

**Kenji Watanabe:** Funding acquisition (equal); Resources (equal).

**Takashi Taniguchi:** Funding acquisition (equal); Resources (equal).

**Tomoki Machida:** Conceptualization (equal); Formal analysis (equal); Funding acquisition (equal); Project administration (equal); Supervision (equal); Validation (equal); Visualization (equal); Writing – review & editing (equal).

## Data availability

The data that support the findings of this study are available from the corresponding authors upon reasonable request.